\documentclass[12pt,fleqn]{article}
\usepackage{amssymb}
\usepackage{hyperref}
\usepackage{amsmath}

\usepackage{amsmath,amssymb,amsthm}

\begin{document}
\begin{center}
{\Large \textbf{Eikonal Equation 0.1}}

\vskip 20pt {\large \textbf{Iryna YEHORCHENKO}}

\vskip 20pt {Institute of Mathematics of NAS Ukraine, 3 Tereshchenkivs'ka Str., Kyiv-4, Ukraine} \\
E-mail: iyegorch@imath.kiev.ua
\\
Institute of Mathematics, PAN, Poland\\
iyehorchenko@impan.pl
\end{center}

\footnotesize
\begin{abstract}
We provide a review of some symmetry-related literature on the eikonal equations
$u_\mu u_\mu =0$,$u_\mu u_\mu =1$,
where lower indices at  dependent variables designate derivatives, $\mu=0,1,2,..,n$ and summation is implied over the repeated indices.

We will consider general solutions and symmetries of these equations, and relations of these equations with reduction of higher-order PDE. Some new results that were needed for comprehensive presentation are also adduced. In particular, we will also consider discrete symmetries of the eikonal equations equivalence classes of solutions and relations of the symmetry solutions and the general solutions. We describe in detail the procedure that allowed obtaining of the general solution using hodograph and contact transformations of the initial equations. Some new results by the author that were needed for comprehensive presentation are also adduced - in particular, a general solution of the eikonal equation for an arbitrary number of space variables.

The paper is intended both as a reference material and as a manual.
\end{abstract}

\normalsize
\section{Notations and Arrangements}
We will consider here a class of nonlinear first-order partial differential equations (PDE) in the following form:
\begin{equation}\label{eik1class}
u_\mu u_\mu =f(x_\mu,u),
\end{equation}
\noindent
where in general we have one real-valued dependent variable $u=u(x_0,x_1,\ldots,x_n)$, a time variable $x_0$ and $n$ space variables $x_1,\ldots,x_n$. I would like to point out that $u$ and other dependent variable take values in a real space, as the complex-valued dependent variables require different treatment and a special study.

If not indicated otherwise, the lower indices of dependent variables designate derivatives with respect to the relevant independent variables $x_\mu$, $\mu=0,1,...,n$. Here we will present results for an arbitrary number of the space variables, but will consider in detail examples for specific cases of
$n=1,2,3$.

If not indicated otherwise, the lowercase Greek letters used as indices are assumed to run from $0$ to $n$, where $n$ is an arbitrary natural number, usually $n>1$ (the case $n=1$ needs special consideration and will be discussed separately. Lowercase Greek letters used as indices (e.g. $\mu$ will designate the $\mu$-th independent variable $x_\mu$ or used as an index at a symbol for a dependent variable the partial derivative with respect to $x_\mu$).

Also if not indicated otherwise, the lowercase Latin letters used as indices are assumed to run from $1$ to $n$, where $n$ is an arbitrary natural number. Usually $n>1$ (the case $n=1$ needs special consideration and will be discussed separately). The lowercase $n$ will designate the number of the independent space variables, or when used as an index will designate the $n$-th independent variable $x_n$ or the partial derivative with respect to $x_n$. Other lowercase Latin letters used as indices (e.g. $k$ will designate the $k$-th independent variable $x_k$ or used as an index at a symbol for a dependent variable the partial derivative with respect to $x_k$).

The following summation arrangements will be assumed (if not specifically indicated otherwise): repeated lowercase Latin indices will mean summation from $1$ to $n$ in the Euclidean space:
 \[
 x_k  x_k = x_1  x_1 + \ldots + x_n  x_n.
 \]
 A summation arrangement using an expression on indices will be used as follows ($m$ takes values from 1 to $n-k$):
 \[
 x_{k+m}  w_m = x_{k+1} w_1 + \ldots + x_n w_{n-k}.
 \]
 The repeated lowercase Greek indices will mean summation from $0$ to $n$ in the Minkowski space:
  \[
 x_\mu  x_\mu = x_0  x_0 - x_1  x_1 - \ldots - x_n  x_n.
 \]
 The indices at dependent variables that are used to indicate partial derivatives will mean the following:
  \begin{gather}\label{deriv}
 u_k=\frac{\partial u}{\partial x_k}, \qquad
 u_{kl} =\frac{\partial u}{\partial x_k\partial x_l},
  \end{gather}
    \noindent
    with a similar formulae applicable to the lowercase
  Greek indices at dependent variables, and to higher-order partial derivatives.

  A letter with an index below
  \[
  \mathop{u}\limits_m
  \]
  \noindent
means the set of all partial derivatives of the function $u$ of the order $m$, with respect to all its independent variables.

$g_{\mu \nu}$ is a metric tensor
$$g_{\mu \nu}={\rm diag}\{1,-1,\ldots,-1\}$$
in an $(n+1)$-dimensional Minkowski space.

$\delta_{ab}$ is a metric tensor
$$\delta_{ab}={\rm diag}\{1,1,\ldots,1\}$$
in an $n$-dimensional Euclidean space.

We will use the following notations in presentation of first-order differential operators:
\begin{gather}\nonumber
 \partial_{\mu}=\frac{\partial }{\partial x_{\mu}}, \qquad
 \partial_u=\frac{\partial }{\partial u}, \qquad  \partial_{u_{\mu}}=
 \frac{\partial }{\partial{u_{\mu}}}.
\end{gather}
  We assume that all functions considered are sufficiently smooth, and all dependent and independent variables take values in the real space.

There is a serious problem with equation names in the mathematics in general. The name "eikonal equation" might be used for something different than
(\ref{eik1class}), and the equations from the class (\ref{eik1class}) or with specific values of $f(x_\mu,u)$ may be called also "Hamilton equations" for $f(x_\mu,u)=1$. E.g. in \cite{Boyer1978} the same equation (\ref{eik1class}) for $f=0$, $n=2$ was called a Hamilton-Jacobi equation.

We will call the Hamilton-Jacobi equation the following
equation:
\begin{equation}\label{HJ}
u_t - u_a u_a =f(t,x_a,u).
\end{equation}

\section{Introduction}
The idea of this paper emerged from the need of the author to have a comprehensive reference material to be cited in the papers presenting new results, and from the lack of anything suitable in the existing literature. It looks like many useful formulae from the author's manuscript notebooks and not actually accessible literature, e.g. \cite{FZhR-preprint}, are worth to be published for the mathematical community in general. It is worth noting that even a Wikipedia article "Eikonal Equation" \cite{wiki} as of 30.11.2022 does not contain any information on the symmetry and on the known parametric general solution of the eikonal equation for low dimensions.

The intended audience of this paper is wider than "the symmetry analysis people", so many detailed arrangements and calculations that are obvious for the above group are written explicitly. The results that can be attributed to other authors are cited as such. However, I failed to find relevant citations for many results and calculations, even though they are not difficult and are probably of a student-exercise level, but needed in a reference paper. Such results will be presented without citations, and I will greatly appreciate any information for the relevant references if available. The list of the results in this paper I believe at the moment of writing to be new will be adduced at the end. The literature cited cannot be regarded as a comprehensive list, I mentioned only was immediately relevant for the presentation.

\section{Symmetry of the eikonal equations with 0 and 1}
Full group classification for the equations of the form
\begin{equation}\label{eik-gen}
  u_au_a = F(t, u, u_t)
\end{equation}
was performed in \cite{preprintIY-RP2001}.

There are two eikonal equations, each having remarkable symmetry properties:
\begin{equation} \label{eik0}
u_\mu u_\mu =0,
\end{equation}
\noindent
and
\begin{equation}\label{eik1}
u_\mu u_\mu =1.
\end{equation}
The maximal Lie symmetry algebra of continuous transformations for the equation (\ref{eik0}) may be described by the following basis operators:
\begin{gather} \label{symmeik0}
  X=2g_{\mu \nu}c^{\mu}(u) x_{\nu}x_{\beta}\partial_{\beta} -g_{\mu \nu}c^{\beta}(u) x_{\mu}x_{\nu}\partial_{\beta}+
   g_{\mu \nu}b_{\mu \beta}(u)x_{\nu}\partial_{\beta}+\\ \nonumber
  +d(u)x_{\beta}(u)\partial_{\beta}+a^{\beta}(u)\partial_{\beta}+
  \eta(u)\partial_u;
  \end{gather}
\noindent
and here $c^{\mu}$, $b_{\mu \beta}(u)$, $d(u)$, and $a^{\beta}$ are arbitrary suf\/f\/iciently smooth functions.

The maximal symmetry algebra of continuous transformations for the equation (\ref{eik1}) may be described by the following basis operators:
\begin{gather}\label{symmeik1}
  \partial_{\mu},\qquad \partial_u, \qquad J_{ab}=x_a\partial_b - x_b\partial_a, \\    \nonumber
  J_{0a}=x_0\partial_b + x_b\partial_0,\qquad
  J_{u0}=u \partial_0 + x_0\partial_u,\qquad
  J_{ua}=u\partial_a - x_a\partial_u,\\    \nonumber
  D=x_0\partial_0 +x_a\partial_a +u\partial_u, \qquad
  K_a=2x_aD+s^2\partial_a,\qquad K_u=2uD+s^2\partial_u, \\    \nonumber
  K_0=2x_0D-s^2\partial_0 \qquad (s^2=x_{\mu}x_{\mu}-u^2).
\end{gather}
Here we used the forms of the symmetry operators listed in \cite{preprintIY-RP2001}. In (\ref{symmeik0}), (\ref{symmeik1}) we assume that $n$ is arbitrary, though to our knowledge these results were strictky proved only for the case $n=3$.

The maximal Lie symmetry of the equation (\ref{symmeik1}) for $n=3$ was found in \cite{FSSht1982}, and for (\ref{symmeik0}) where $n=3$ it was found in \cite{FSerovLdAE}. Lie symmetry for (\ref{symmeik0}), $n=2$ was described in \cite{Boyer1978}.
There exist some papers with the titles like "Symmetries of the Eikonal equation" (see e.g. \cite{Ghanam2018}) where authors are obviously not acquainted with the previous literature on the subject and basics of the Lie symmetry analysis, do not cite many available papers on low-dimensional eikonal equations and either repeat previously known results or present not substantiated and not comprehensive results.

So, I see the need to give here a proof of the results on the maximal Lie symmetry of the equations (\ref{eik0}) and (\ref{eik1}) stated by the formulae in (\ref{symmeik0}), (\ref{symmeik1}).

A proof would require going to the basics of the Lie symmetry analysis procedures (see e.g. \cite{Olver1}).

An equation $F(x,u,\underset{1}{u},\ldots , \mathop{u}\limits_k)=0$ (we consider here one scalar functions) is
invariant under a first-order differential operator
\begin{gather} \label{X}
X=\xi^{\mu}(x_{\nu},u)\partial_{x_{\mu}}+
\eta(x_{\nu},u)\partial_{u}
\end{gather}
\noindent
such that
\begin{equation}\label{cond-ofinv}
  \mathop{X}\limits_k F \big |_{F=0}=0,
\end{equation}
\noindent
where $\mathop{X}\limits_k$ is a $k$-th prolongation of the operator $X$.

A proof that some set of first-order differential operators is indeed a maximal Lie invariance algebra of the equation being considered is to prove that a linear combination of the operators from this set is a maximal operator satisfying the condition (\ref{cond-ofinv}).

For our particular proof we need only first prolongation of the operator $X$ (\ref{X}) (this is the operator acting not only on independent and dependent variables, but also at the derivatives, and its coefficients are fully determined by the coefficients of the operator $X$:
\begin{gather} \label{X1}
 \mathop{X}\limits_1=X+\zeta^{\mu}(x_{\mu},u)\partial_{u_{\mu}},
\end{gather}
\noindent
where the coefficients
\begin{gather} \label{zeta}
 \zeta^{\mu}(x_{\mu},u)=\eta_{\mu}+\eta_u u_\mu- \xi^{\nu}_{\mu}u_{\nu} -\xi^{\nu}_uu_{\mu}u_{\nu}.
\end{gather}
Application of the operator (\ref{X1}),(\ref{zeta})  to the equations
(\ref{eik0}) and (\ref{eik1}) gives the following expression:
\begin{gather} \label{X1eik}
 \mathop{X}\limits_1 u_{\mu} u_{\mu}= 2\zeta^{\mu}u_{\mu} = 2(\eta_{\mu}u_{\mu} +\eta_u u_{\mu} u_{\mu} - \xi^{\nu}_{\mu}u_{\nu} u_{\mu} -\xi^{\nu}_uu_{\mu}u_{\mu}u_{\nu})= \\ \nonumber
 2(\eta_{\mu}u_{\mu} - \xi^{\nu}_{\mu}u_{\nu} u_{\mu}) +
 2u_{\mu} u_{\mu}(\eta_u -\xi^{\nu}_uu_{\nu})
\end{gather}
From the condition that (\ref{X1eik}) has to vanish (even before transition to the relevant manifolds determined by equations
(\ref{eik0}) and (\ref{eik1})) we get for both equations
the following conditions for the coefficients of the operator $X$ from coefficients at $u_{\mu}u_{\nu}$:
\begin{gather} \label{inv1}
\xi^{\nu}_{\mu}+\xi^{\nu}_{\mu}=0 \\
\xi^{\nu}_{\nu}=\xi^{\mu}_{\mu} \qquad {\rm if} \qquad \nu \ne \mu. \nonumber
\end{gather}

No summation over the repeated indices is implied in (\ref{inv1}).

Transition to the manifold of equation (\ref{eik0})
gives more conditions in addition to (\ref{inv1}):
\begin{gather} \label{inv2}
\eta_{\mu} =0.
\end{gather}
From (\ref{inv1}) and (\ref{inv2}) we get explicit forms of coefficients of the operator $X$ (\ref{symmeik0}).

Transition to the manifold of equation (\ref{eik1})
gives more conditions in addition to (\ref{inv1}):
\begin{gather} \label{inv3}
\eta_{\mu} = \xi^{\mu}_u, \qquad \eta_u = \xi^{\mu}_{\mu},
\end{gather}
\noindent
no summation over the repeated indices is implied in (\ref{inv3}).

From the equations (\ref{inv1}) and (\ref{inv3}) we get coefficients of the operator $X$  and the basis of the maximal symmetry algebra of continuous transformations (\ref{symmeik1}). Conditions (\ref{inv1}) imply presence of the Poincar\'e and conformal algebras. For (\ref{eik1}) these algebras appear in the space of one more variable than e.g. for the wave equation
$\Box u=0$ in $n$ space dimensions - the dependent variable is included in the same way as the independent variables. The maximal Lie invariance algebra for (\ref{eik0}) appears to be infinitely dimensional, and coefficients of its operators include arbitrary functions of the dependent variable $u$. It is easy to check that (\ref{eik0}) has a remarkable property - a function of any of its solutions is also a solution.

The discrete transformations for the equations (\ref{eik0}) and (\ref{eik1}) are quite obvious. But for a comprehensive presentation we need to note that these equations are invariant under time reflection
$t'=-t$, space reflections $x'_{\mu}=-x_{\mu}$, dependent variable reflection $u'=-u$, under a permutation group for the independent space variables. A hodograph transformation $x'_1=u$, $u'=x_1$ that is a local transformation for (\ref{eik1}) will also be a discrete transformation for this equation.

\section{General Solutions}
In consideration of the general solutions for the eikonal equations, we will follow the algorithm, given in \cite{FZhR-preprint}, where general solutions for the equations (\ref{eik1}) for $n=3$ are adduced. We generalize the relevant formulae for an arbitrary number of space variables.

The algorithm in \cite{FZhR-preprint} uses contact transformations, nonlocal transformations involving dependent and independent variables and first derivatives. The general form of such contact transformations is as follows:
\begin{gather} \nonumber
  x'_{\mu}= \varphi_{\mu}(x_{\mu},u,u_{\mu})\\ \label{contactgen}
  u' = \Phi(x_{\mu},u,u_{\mu})\\ \nonumber
  u'_{\mu} =\Phi_{\mu}(x_{\mu},u,u_{\mu}).
\end{gather}
Here $x'_{\mu}, u'$ are new transformed variables,
 $u'_{\mu}$ are transformed first derivatives, $\varphi_{\mu}$, $\Phi$ and $\Phi_{\mu}$ are sufficiently smooth functions on their variables. Note that indices at $\Phi_{\mu}$ are just indices, not derivatives.

 We will first present in detail the procedure for construction of the general solution for the simplest PDE of the type (\ref{eik1}) for only two space variables:
 \begin{equation}\label{eik1-2}
u_1^2+u_2^2=1.
\end{equation}
Note that if we apply the partial derivative operators to  (\ref{eik1-2}), we get the system
\begin{gather} \label{2deriveik1-2}
 u_1u_{11}+u_2u_{12}=0,\\
 u_1u_{21}+u_2u_{22}=0,
\end{gather}
\noindent
from which we can see linear dependence of the above equations considered as linear equations for the first derivatives of $u$, $u_1$ and $u_2$.

So, the determinant of the matrix of the second derivatives of $u$
\begin{gather}
U={\left(
\begin{matrix}
    u_{11} & u_{12} \nonumber \\
    u_{21} & u_{22} \nonumber \\
  \end{matrix}
  \right)}
\end{gather}
\noindent
is equal to zero. However, here there will be two cases to be treated separately:
\begin{itemize}
  \item ${\rm rank}(U) =0$,
  \item ${\rm rank}(U) =1$.
\end{itemize}

In the first case, we have $u_1=u_2=const$, and $u$ linearly dependent on $x_1$ and $x_2$:
\begin{equation}\label{gseik1-2lin}
 u=c_1x_1+c_2x_2+c,
\end{equation}
\noindent
where $c_1$, $c_2$ and $c$ are arbitrary constants such that $c_1^2+c_2^2=1$.

In the second case we obtain an implicit general solution depending on a parameter by means of contact transformations.

We apply to (\ref{eik1-2}) the following contact transformations in the case if $u_{x_1}\ne 0$:
\begin{gather} \nonumber
  H=x_1 u_{x_1}-u, \qquad u_{x_1}=y_1, \qquad x_2=y_2;\\ \label{contacteik2v}
  H_{y_1}=x_1, \\  \nonumber
  H_{y_2}=-u_{x_2};
\end{gather}
\noindent
and respectively
\begin{gather} \nonumber
  u=y_1 H_{y_1}-H, \qquad H_{y_1}=x_1, \qquad y_2=x_2;\\ \label{contacteik2w}
  u_{x_1}=y_1, \\ \nonumber
  u_{x_2}=-H_{y_2}.
\end{gather}
The resulting equation is a linearisable ordinary differential equation (ODE)
\begin{equation}\label{eik1-2cont1}
H^2_{y_2}+y^2_1=1.
\end{equation}
We can linearize it as follows:
\begin{equation}\label{eik1-2cont2}
H_{y_2}=\sqrt{1-y^2_1}
\end{equation}
A general solution of (\ref{eik1-2cont2}) may be taken as follows:
\begin{equation}\label{gseik1-2cont3}
H=-y_2\sqrt{1-y^2_1}-\Psi(y_1),
\end{equation}
\noindent
where $\Psi(y_1)$ is a sufficiently smooth arbitrary function on $y_1$. Further it is convenient to introduce a parameter $\tau$ to be used instead of $y_1$.

Transformations (\ref{contacteik2w}) allow going back to expressions for the function $u$, and we obtain the following general solution for (\ref{eik1-2}):
\begin{gather} \label{gseik1-2a}
  u=\tau H_{\tau}-H =x_1\tau+x_2\sqrt{1-\tau^2}+\Psi(\tau), \\ \nonumber
   x_1=H_{\tau}= x_2(\sqrt{1-\tau^2})^{-1}-\Psi_{\tau}(\tau),\nonumber
 \end{gather}
\noindent
where $\Psi(\tau)$ is a sufficiently smooth arbitrary function on $\tau$.

We may rewrite the expressions (\ref{gseik1-2a}) in a more convenient form:
\begin{gather} \label{gseik1-2b}
  u=x_1\tau+x_2\sqrt{1-\tau^2}+\Psi(\tau), \\ \nonumber
   x_1-x_2(\sqrt{1-\tau^2})^{-1}+\Psi_{\tau}(\tau) =0\nonumber
 \end{gather}
In our calculations that used the contact transformations, we assumed  $u_{x_1}\ne 0$, or, more generally, the rank of the matrix of the second derivatives of the function $u$
\begin{gather}
{\left(
\begin{matrix}
    u_{11} & u_{12} \nonumber \\
    u_{21} & u_{22} \nonumber \\
  \end{matrix}
  \right)}
\end{gather}
\noindent
is equal to one.

Let us consider the case of equation (\ref{eik1})
with an arbitrary number of space variables.

Similarly to (\ref{2deriveik1-2}), we can apply the partial derivative operators to (\ref{eik1}) and conclude linear dependence of the resulting equations considered as linear equations for the first derivatives of $u$, and that the rank of the matrix of the second derivatives of the function $u$ may take values from 0 to $n$.

In the case when the rank of the matrix of the second derivatives of the function $u$ is equal to 0,
all first derivatives $u_{\mu}$ are constants, and $u$ is linearly dependent on $x_{\mu}$:
\begin{equation}\label{gseik1-nlin}
 u=c_{\mu}x_{\mu}+c,
\end{equation}
\noindent
where $c_{\mu}$ and $c$ are arbitrary constants such that $c_{\mu}c_{\mu}=1$.

In the case when the rank of the matrix of the second derivatives (we will use the notation $U$ for this matrix) of the function $u$ is equal to $n$, we can apply a procedure very similar to what was used above for the equation (\ref{eik1-2}), and by means of the contact transformations (see \cite{FZhR-preprint} where it is considered for $n=3$)
\begin{gather} \nonumber
  H=x_a u_{x_a}-u, \qquad y_a=u_{x_a}, \qquad x_0=y_0;\\ \label{contacteik1n}
  H_{y_a}=x_a, \\  \nonumber
  H_{y_0}=-u_{x_0};
 \end{gather}
\noindent
obtain an $n$-parameter general solution (${\rm rank}(U) =n$):
\begin{gather} \label{gseik1-nn}
  u=-x_a\tau_a+x_0\sqrt{1+\tau_d\tau_d}+\Psi(\tau_d), \\ \nonumber
   x_a-x_0\tau_a(\sqrt{1+\tau_d\tau_d})^{-1}-\Psi_{\tau_d}(\tau_a) =0 \nonumber
 \end{gather}
\noindent
where $\Psi(\tau_a)$ is a sufficiently smooth arbitrary function on $n$ parameters $\tau_a$.

Further general solutions for every rank $k=1,n-1$ will be written depending on $k$ parameters.

In each such case we may introduce $n-k$ new functions $w_m$, were $m$  are indices taking values from 1 to $n-k$. Without any limitation of generality, we can assume
\begin{gather}\label{w}
  u_0=\sqrt{1+u_bu_b+w_mw_m}, \\ \nonumber
  u_{k+m}=w_m(u_b), \nonumber
\end{gather}
\noindent
where the indices $m$ run from 1 to $n-k$, and the indices and symbols of derivatives $b$ run from 1 to $k$, symbols of derivatives $k+m$ run from $k+1$ to $n$. Thus, we can take for every rank $k=1,n-1$ of the matrix $U$ $n-k$ new functions $u_{k+m}=w_m(u_b)$ depending on the first $k$ derivatives $u_b$.
General solutions depending on $k$ parameters may be taken as follows:
\begin{gather} \label{gseik1-nk}
  u=-x_b\tau_b+x_0\sqrt{1+\tau_d\tau_d+w_mw_m}+w_mx_{k+m}
  +\Psi(\tau_d), \\ \nonumber
   x_b-x_0(\sqrt{1+\tau_d\tau_d+w_mw_m})^{-1}
   -w_{m\tau_b}x_{k+m}-\Psi_{\tau_b}(\tau_d) =0
 \end{gather}

The expressions in (\ref{gseik1-nk}) imply a maybe unusual summation over $m$ from 1 to $n-k$.

Proof of the fact that (\ref{gseik1-nk}) is a rank $k$ general solution of (\ref{eik1}) for $n$ space variables can be done by means of a contact transformation
\begin{gather} \nonumber
  H=x_b u_{x_b}-u, \qquad y_{k+m}=u_{x_{k+m}}, \qquad x_0=y_0; \qquad x_b=y_b;\\ \label{contacteik1-n}
  H_{y_b}=x_b
  H_{y_0}=-u_{x_0}, \\  \nonumber
  H_{y_{k+m}}=-u_{x_{k+m}}
 \end{gather}
\noindent
applied similarly to the procedures in \cite{FZhR-preprint}.

A general solution for (\ref{eik0}) and for arbitrary number of independent variables was found in \cite{Collins1}.

\section{A Note of the General Solution and the Known Symmetry Exact Solutions for Eikonal Equations}
A well-known symmetry solution - the radial solution of the equation (\ref{eik1})
\begin{equation}\label{rad1}
u=\sqrt{x_{\mu}x_{\mu}}
\end{equation}
\noindent
has the rank $n$ and corresponds to the following solution represented by the general solution formula (\ref{gseik1-nn})
\begin{gather} \label{gseik1-rad1}
  u=x_a\tau_a-x_0\sqrt{1+\tau_b\tau_b}, \\ \nonumber
   x_a-x_0\tau_a(\sqrt{1+\tau_b\tau_b})^{-1}=0,
 \end{gather}
The parameters $\tau_a$ will take values
\begin{equation} \label{taurad}
  \tau_a=\frac{x_a}{\sqrt{x_{\mu}x_{\mu}}}.
 \end{equation}
The formula (\ref{gseik1-nn}) includes arbitrary functions, so all solutions (\ref{gseik1-nn}) cannot be equivalent to symmetry solutions. In general (not always), formulas for the general solutions give wider classes of solutions than the Lie symmetry method.

As to the zero-rank general solution (\ref{gseik1-nlin}), all such solutions are symmetry solutions. For more symmetry solutions of low-dimensional eikonal equations see e.g. \cite{FSerovLdAE} and \cite{VFVF2016}.

\section{Some Relations Among the Eikonal Equations and
the Hamilton-Jacobi Equation}
As was established by Sophus Lie \cite{Lie1}, all scalar first-order partial differential equations can be transformed into one another by some suitable contact transformations.

A hodograph transformation
\begin{equation} \label{hd}
u=y_0, \qquad x_0=w,\qquad  x_a=y_a
\end{equation}
will get from a system of two eikonal equations
\begin{gather}
u_\mu u_\mu =0,\\ \nonumber
v_\mu v_\mu =0, \label{eik2}
\end{gather}
a system of an eikonal equation without a time variable (with only $n$ space variables) and a Hamilton-Jacobi equation
\begin{gather}
w_{y_a}w_{y_a}=1, \nonumber \\
v_{y_a}v_{y_a}=2v_{y_0},  \label{eikHJ}
\end{gather}

\section{New Results, Conclusions and Further Work}
The new results of this paper are a proof of maximal Lie symmetries for the eikonal equations (\ref{eik0}) and (\ref{eik1}) and a general solution for (\ref{eik1}) for arbitrary number of space variables.

The eikonal equation is important in many fields, and in symmetry analysis it often appears in reduction conditions for many higher-order equations, in particular, for all equations invariant under the orthogonal and Poincar\'e groups (\cite{FZhR-JMPh}, \cite{anzSch}). Its general solution is a first step to find general solutions of reduction conditions.

There are also multiple papers searching for numerical solutions of the boundary problems for eikonal equations. On my opinion, it would be interesting to look for these using the general solutions.

Further work will include study of the coupled eikonal equations in higher dimensions (for the case of 2 space variables see \cite{preprintIY2017}), relevant conditional and hidden symmetries and looking for higher-order equations with the same or similar symmetry properties as the eikonal equation.

\section{Acknowledgements}
The first and foremost my acknowledgements go to the
Armed Forces of Ukraine and to the Territorial Defence Forces of Ukrainian Regions due to whom I am alive and is able to work.

Please remember that Russia is an aggressor country and still plans to kill all Ukrainians not going to be their subordinates, and Russian scientists contribute to killings despite words about peace from a tiny portion of them; if even by trying to persuade na\"{i}ve people that Russia is a civilised country.

I would like also to thank the Institute of Mathematics of the Polish Academy of Sciences for their hospitality and grant support, to the National Academy of Sciences of the USA and the National Centre of Science of Poland for their grant support.

Research was supported by Narodowe Centrum Nauki, grant number 2017/26/A/ST1/00189.

\end{document}